# The Recurrent Nova TCrB: A Method for Predicting the Next Eruptive Event in Nova Cycles

**by Stefano Sello (Mathematical and Physical Models, Enel Research, Pisa – Italy)**

**email: deneb50cyg@libero.it**

*Abstract*

The symbiotic recurrent nova (SyRNe) TCrB (T-Coronae Borealis) is perhaps the most famous example of the group of known four symbiotic nova systems, for which at least two previous nova eruptions are known and accurately recorded: in 1866 and 1946. B.E. Schaefer (2023) has identified the dates of two other previous eruptive events: in 1787 and 1217. Its peak magnitude V was found to be 2.50 ± 0.10, making it the brightest of its class. In its quiescent phase, TCrB is the brightest of all known novae, with a mean magnitude of 9.8. Careful studies, especially photometric ones, have led to different predictions for the next nova eruption, taking into account the recurrence times extrapolated from previous eruptions, which an average value about 80 years. Schaefer, in particular, has produced various forecasts, including one made in 2023 based on B&V light curves for the period: 1842–2022, which predicts the next nova eruption should occur in 2025.5±1.3 and is therefore still valid today. Using the Schaefer's remarkable work in accurately determining the key physical parameters that drive the dynamics of the TCrB symbiotic system, we propose here a new semi-empirical method to derive the variations in the nova recurrence time, Trec, and thus obtain a forecast estimate for the next eruption for the date: 26-Feb-2027, which is currently compatible and consistent with the observed behavior and would also justify the supposed "delay" for the next event of this nova as commented by various authors.

## 1. *Introduction*

Recurrent novae represent a rather heterogeneous set of a small group of similar objects within the larger class of cataclysmic variable systems, CV (Payne-Gaposchkin (1957), Warner, (1995)). All cataclysmic variables are made up of close binary systems, where significant mass transfers occur between the components. Recurrent novae, in particular, are classical nova events, powered by thermonuclear reactions, which recur on relatively short timescales, < 100 years. These recurrence times, Trec, distinguish them from dwarf novae, which are much shorter, and from classical novae, which are much longer and for which only one eruptive event has been observed and recorded.

Here we adopt the criteria provided by Webbink et al., (1987), to well define a recurrent nova: 1) two or more distinct eruptive events recorded, with peak absolute magnitude comparable to those of classical novae: Mv<~ -5.5); 2) ejection of a significant shell of matter during the nova event with velocities comparable to those of classical novae: $v_{exp}$>300 km $s^{-1}$; 3) Mass transfer must occur from an expanded companion cold star onto a dense degenerate star, a white dwarf (WD), of mass close to the Chandrasekhar limit (~1.43 $M_{\odot}$).

In particular, the recurrent nova TCrB (T-Coronae Borealis) is also defined as a symbiotic system (SyRNe), (Schaefer, 2025b) characterized by a red giant companion star with an intense and dense stellar wind that feeds the white dwarf.

## 2. *Main properties of the SyRNe TCrB System*

The main properties of TCrb are listed in Table 1. Recall that the parameter t3 indicates the time in days required for a decrease of 3 magnitudes from the peak magnitude. From the table, it is clear that TCrb is a fast and a long-period nova.

| Name | Type | $M_V$ peak | t3 (days) | Dist (kpc) | Porb (days) | $v_{ejecta}$ (km $s^{-1}$) | Sp (comp) | $M_V$ peak |
|---|---|---|---|---|---|---|---|---|
| TCrB | SyRNe | 2.50± 0.10 | 6 | 0.914 | 227.532 | 4980 | M4 III | -7.62 |

Tab.1 – Main properties of TCrB (Schaefer, (2022)).

In recent publications dated 2025, Schaefer addresses the problem of verifying whether the TCrB system can give rise to a type Ia supernova in the future in its evolution. This, the question of the possible progenitors of type Ia supernovae, is a topic of extreme interest in various fields of astrophysics. Using different approaches, Schaefer demonstrates, with great clarity and accuracy, that TCrB will not be able to produce a type Ia supernova (see for full details: Schaefer (2025a), (2025b).

Here we are particularly interested in the analysis made by Schaefer to estimate the slow and rapid variations of the orbital period, Porb, in the intervals between two nova eruptions and during the eruption, using a large database of radial velocities in the period: 1946-2025 and photometric data in the period: 1866-2025, Schaefer (2025a).

For most of the ~80 year interval between two successive eruptions, TCrB is in a low-activity state with a mass accretion rate of: $dM/dt_{low}=3.2\times10^{-9}\ M_\odot\ yr^{-1}$. In the two decades before and after each nova eruption, TCrB is in a high-activity state with a mass accretion rate of: $dM/dt_{high}=6.4\times10^{-8}\ M_\odot\ yr^{-1}$, (Schaefer (2014)).

Table 2 lists the main results obtained from the careful analysis by Schaefer (2025a). $M_{WD}$ = current white dwarf mass; Ppre = orbital period before the 1946 eruption; Ppost = orbital period after the 1946 eruption; ΔP = Ppost-Pre; dP/dt = slow variation between two successive eruptions; Mejecta = mass ejected during the 1946 nova eruption; Maccreted = mass accreted via mass transfer during the two subsequent eruptions; dM/dt = rate of mass transfer from the companion to the white dwarf (high state).

| $M_{WD}$ | Mcomp | Ppre (days) | Ppost (days) | ΔP=Ppost-Ppre (quick var.) er.1946 | dP/dt (slow var. between two eruptions) | Mejecta (Kepler law) er.1946 | Maccreted | dM/dt |
|---|---|---|---|---|---|---|---|---|
| 1.32± 0.10 $M_\odot$ | 0.98±0.31$M_\odot$ | 227.4586 | 227.6043 | +0.146±0.019 | $-3.1\pm1.6\times10^{-6}$ | $7.4\times10^{-4}\pm0.00009\ M_\odot$ | $1.38(1.47)\times10^{-6}\ M_\odot$ | $6.4\times10^{-8}\ M_\odot\ yr^{-1}$ |

Tab. 2 – Results of the TCrB period variation analysis (Schaefer, 2025a).

The most important result of Schaefer's work (2025a), is the accurate estimate of the mass ejected during the 1946 eruption which has to be compared with the mass accreted during two subsequent eruptions: the remarkable result is that the ratio between the mass ejected during the nova eruption and the mass accreted in the interval between two eruptions is: 540±65. This is strong evidence that TCrB cannot produce a type Ia supernova in the future, being unable to reach the Chandrasekhar mass limit, but rather will progressively consume itself over time.

### 3. *Predictions for the next TCrB nova eruption event*

Several predictions about the next nova event of TCrB have been proposed in the literature after the last recorded one in 1946. L. Peltier in 1945 identified a significant and rapid weakening, months before the main nova eruption and this was interpreted by many as a precursor signal of the incoming eruption. With the discovery of a high mass transfer state and a deep weakening in pre-eruption conditions, suggested to many authors the possibility of making accurate predictions of subsequent eruptions (Schaefer 2014). In particular, in the year 2015 the American Association of Variable Star Observers (AAVSO) recorded a rapid transition to a high state in the light curve in the $B$ band, with a shape similar to that recorded in 1938. This transition, first identified photometrically and spectroscopically by Munari, Dallaporta, & Cherini (2016), with the assumption that the pre-eruption behavior is similar for each eruption, led Schaefer (2019) to predict the next nova eruption at: 2023.6±1.0. The following year Luna et al. (2020) predicted the eruption in the period: 2023–2026. To refine these predictions, Schaefer in 2023 accurately reconstructed a complete light curve, in the modern Johnson $B$ and $V$ systems, over the period: 1842–2022 (see Fig. 3 of Schafer (2023). Considering all the uncertainties involved in the numerical estimates and the similarities between the light curves in subsequent eruptions, Schaefer's best prediction for the post-1946 eruption is: 2025.5±1.3, i.e. from 2024.2 to 2026.8. At the time of writing this article, this prediction is still valid even though we are close to the upper limit of the prediction.

More recently, Schneider (2024) published a study to further refine the prediction of the date of the next eruption of nova TCrB. This work is based on the combination of known past eruption dates (Trec values) and the orbital parameters of the binary system, without making any assumptions about the physical mechanism of the eruptions or the characteristics of the light curves. His conclusion is that the next eruption should occur around March 27 or November 10, 2025, or later. More precisely, starting from the known eruption dates we have:

- 1217 nov 04 JD=2165874.5
- 1787 dic 20 JD=2374101.5 — diff. = 208227 days (570.09 yr)
  Trec ~80 yr, 7 eruptions (81.44 yr)
- 1866 mag 12 JD=2402733.5 — diff.= 28632 days (78.39 yr)
- 1946 feb 09 JD=2431860.5 — diff. = 29127 days (79.74 yr)

and therefore we make the hypothesis that the next eruption can be predicted only on an empirical-numerical basis from the relation: Trec = integer multiple N of the orbital period Porb=227.5687:

(2374101.5-2165874.5)/227.5687 = 915.007 (N=131)
(2402733.5-2374101.5)/227.5687=125.82 (N=126)
(2431860.5-2402733.5)/227.5687=127.99 (N=128)

or equivalently:

Teruption=JD(nova prec)+NxPorb=2431860.5+Nx227.5687 with: N€[126,131]

The possible predicted dates by this method are therefore:

| N value | Teruption prediction |
|---|---|
| 126 | 2460534.16 **2024-Aug-11** |
| 127 | 2460761.72 **2025-Mar-27** |
| 128 | 2460989.29 **2025-Nov-09** |
| 129 | 2461216.86 **2026-Jun-25** |
| 130 | 2461444.43 **2027-Feb-07** |
| 131 | 2461672 **2027-Sep-23** |

with Porb uncertainty of ± 8 days.

At the moment, we can exclude the first four dates that have already passed and verify the last two in the table, with N > 129.

Possible causes for N to vary from eruption to eruption are: a) Time variable rate of accretion of matter both onto the disk and white dwarf surface; b) External sources of perturbation in the eruption mechanism; c) Variations in the orbital eccentricity, etc. The main limit of this method is that it is not possible to predict the actual value of N for the Trec, and therefore all plausible N values are equally probable.

In the next section, we will describe a new method for making an alternative, semi-empirical prediction for the date of the upcoming TCrB eruption.

## 4. *New method for predicting the next TCrB nova eruption event*

The method we will describe in the following, of a semi-empirical nature, is based on all the information obtained and described in Section 2, especially those relating to the mass transfer rates, the variations of the orbital period, Porb, and of the mass ejected during the 1946 eruption, Mejecta, provided by the works of Schaefer (2025a).

Livio (1988) shows that the Trec for a recurrent nova, which as we have seen is much shorter than that of a classical nova, is a suitable function of some important physical parameters: 1) Mwd (white dwarf mass), 2) dM/dt (accretion rate), 3) Lwd (white dwarf luminosity), 4) Z (heavy element content of the accreting material), 5) orbital eccentricity. Considering, as a first approximation, only the first two contributions as significant, we obtain the following formula: (see eq. 2 in Livio (1988))

$$Trec \cong 1.2x10^4 \left(\frac{\frac{dM}{dt}}{10^{-8}M_{\odot}yr^{-1}}\right)^{-1} \left(\frac{M_{WD}}{M_{\odot}}\right)^{-1} \left(\frac{R_{WD}}{6x10^8cm}\right)^4 yr \quad (1)$$

Using a suitable relation for the mass-radius of white dwarfs, $R_{WD}= f(M_{WD})$, one can obtain from (1) a recurrence time, Trec, as a function of the white dwarf mass alone and the accretion rate dM/dt.

Assuming the simplified case of non-relativistic degenerate electrons, to express the density of the stellar core of the white dwarf, the theory allows us to write the following relation, (Norton, (2024)):

$$R_{WD} = \left(\frac{R_{\odot}}{74}\right) x \left(\frac{M_{WD}}{M_{\odot}}\right)^{-1/3} \quad (2)$$

For rather high $M_{WD}$ masses, close to the Chandrasekhar limit, eq. (2) is not sufficiently accurate and another formula must be used which provides a more accurate, empirically based, evaluation of the radius of a white dwarf. The derivation of the formula is due to Nauenberg (1972), and gives:

$$R_{WD} \approx 7.8x10^{8} cm\ x \left[\left(\frac{1.4M_{\odot}}{M_{\mathrm{WD}}}\right)^{2/3} - \left(\frac{M_{\mathrm{WD}}}{1.4M_{\odot}}\right)^{2/3}\right]^{1/2} \quad (3)$$

From eq. (1) we note how critical is the adequate evaluation of the radius $R_{WD}$ because of the dependence of Trec on its fourth power. Numerical experiments with the characteristic parameters of the TCrB show that eq. (2) leads to a strong overestimation of the Trec, while eq. (3) leads to a strong underestimation of the Trec. It is therefore convenient to use an appropriate combined formula of (2) and (3) with adequate weights for a correct estimation of the $R_{WD}$.

Let us therefore put for $R_{WD}$ in formula (1):

$$R_{WD} = w_1 x R_{WD}(3) + w_2 x R_{WD}(2) \quad (4)$$

where the weights $w_1$ and $w_2$ will have to be experimentally calibrated on the basis of known historical data. Given the temporal variability of the Trec, it is natural to expect that the weights will also be variable over time. To numerically determine the pair $(w_1,w_2)$, we used the numerical estimates of the characteristic parameters of the TCrB listed in Table 2 by Schaefer (2025a). In particular, we will use the values of the current mass $M_{WD}$, the Mejecta mass during the last eruption in 1946 and we will make the safe assumption that this mass has remained constant during the various nova eruptions. We will also use the mass transfer rates of the low and high states to evaluate the overall mass accretion between two successive eruptions. From the assumptions made above, it is convenient to use only the information from the last two recorded eruptions, that of 1946 and that of 1866. For the term: dM/dt in eq.(1), we use the formula: $\frac{dM}{dt} = 0.75\left(\frac{dM}{dt}\right)_{low} + 0.25\left(\frac{dM}{dt}\right)_{high}$, where: $\left(\frac{dM}{dt}\right)_{low}$ $and$ $\left(\frac{dM}{dt}\right)_{high}$ are evaluated, as said, during the last interval pre-eruption, (Schaefer, 2025a). The following Table 3 shows the results of the numerical calibration of the model weights eq. (4)

| Data eruzione | $M_{WD}$ [$M_{\odot}$] | $w_1$ | $w_2$ | Trec [yr] |
|---|---|---|---|---|
| feb 9 1946 | 1.32 + Mejecta | 0.7457 | 0.2543 | 79.74 |
| may 12 1866 | 1.32+2xMejecta | 0.7468 | 0.2532 | 78.39 |

Tab. 3 – Calibration of model weights in eq. (4) for TCrB.

Linearly extrapolating the weight variation for the next nova eruption ($\Delta w=\pm 1.1x10^{-3}$) we will have the following prediction:

| Predicted eruption date | $M_{WD}$ [$M_\odot$] | $w_1$ | $w_2$ | Trec [yr] |
|---|---|---|---|---|
| feb 26 2027 | 1.32 | 0.7446 | 0.2554 | 81.049±0.0902 |

Tab.4 – Weight calculation, Trec prediction, and next nova eruption date. The uncertainty about Trec comes from the uncertainty about the value of Mejecta.

Figures 1 and 2 show the temporal trends of the weights $w_1$ e $w_2$, where two are calculated on the last two recorded eruptions and one is predicted, and of the corresponding Trec/Porb values.

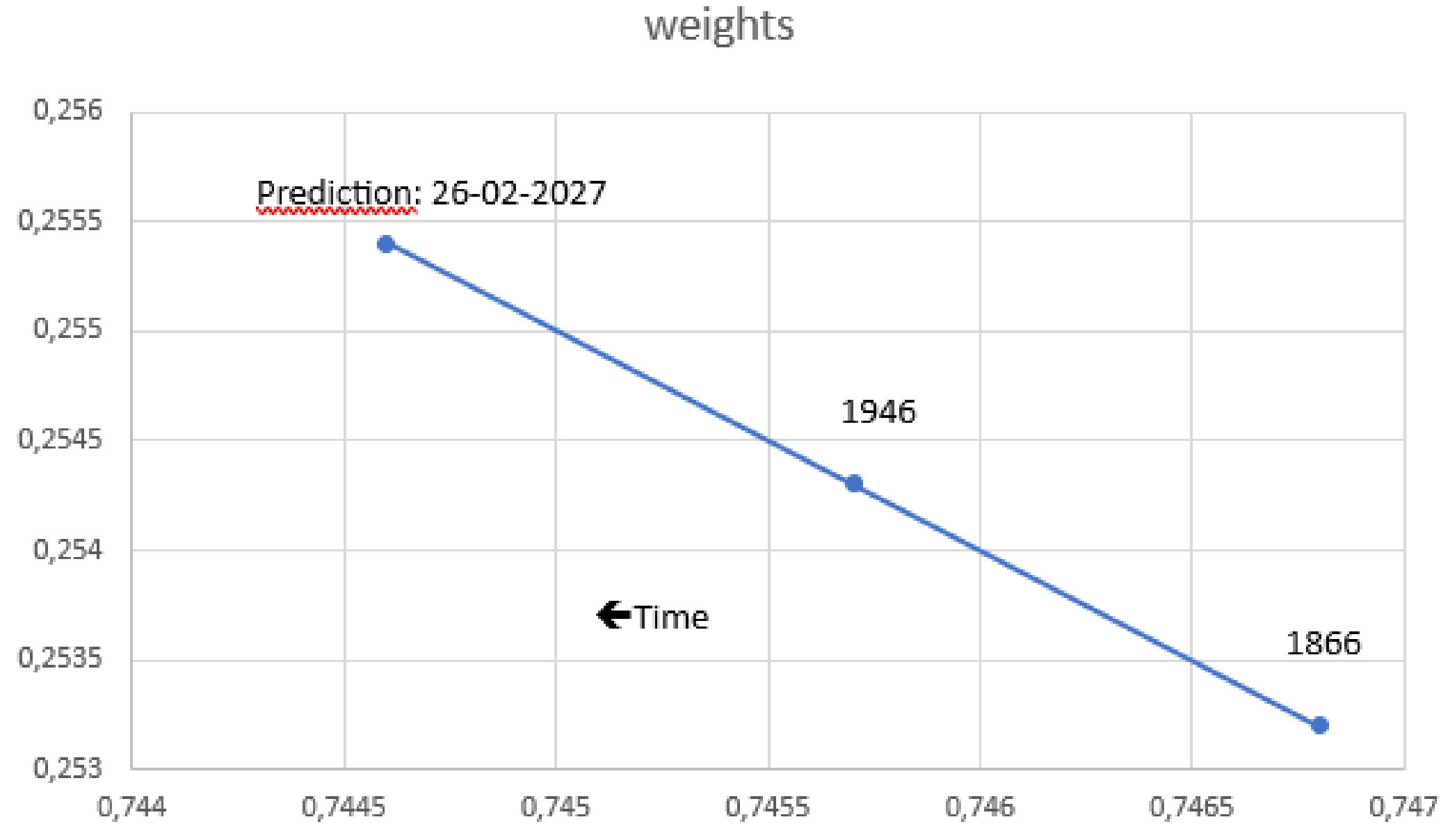

Fig.1 – Weight values ($w_1$,$w_2$) for the latest nova eruptions of TCrB.

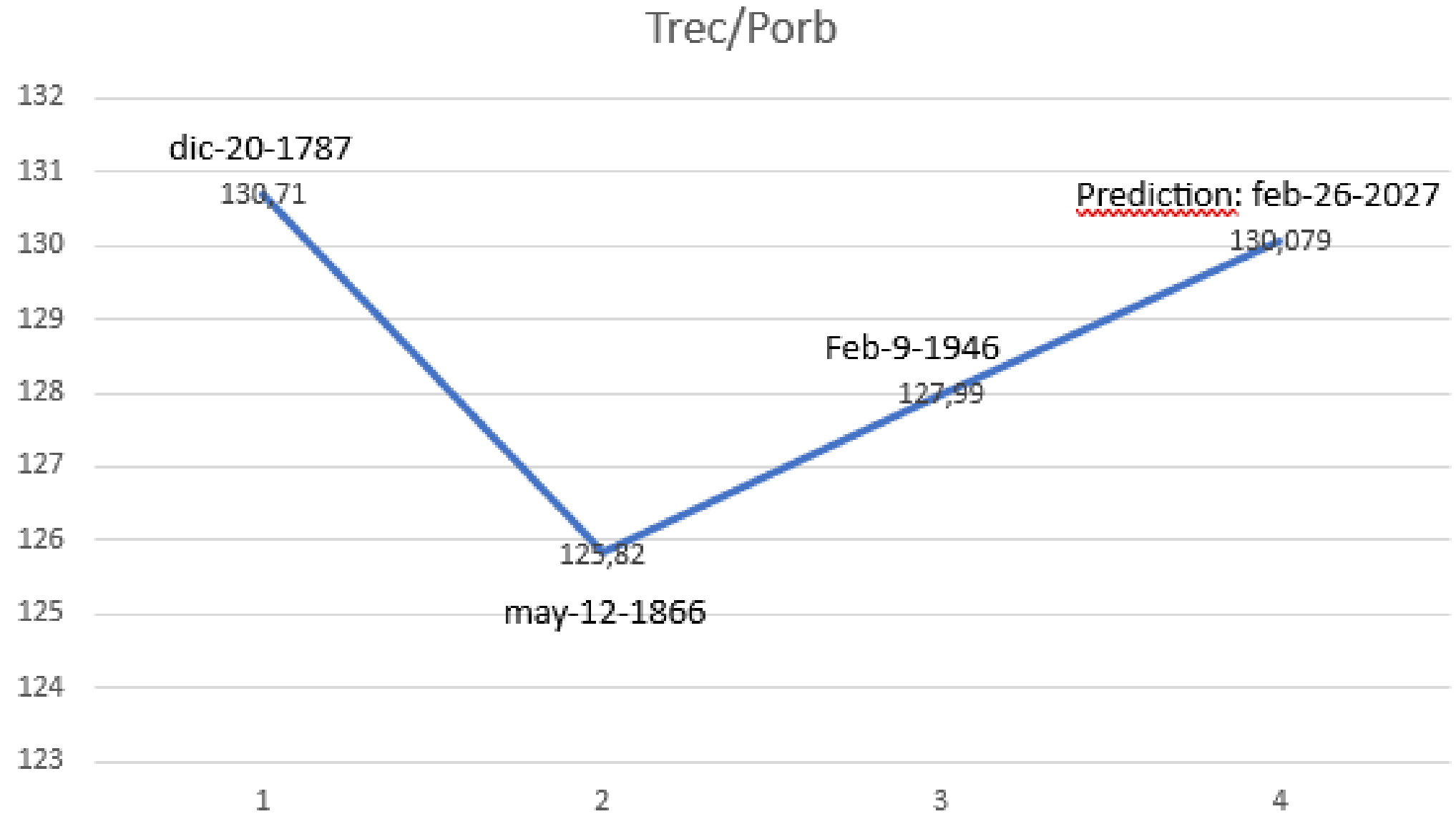

Fig. 2 – Trec/Porb values for the latest nova eruptions of TCrB. (Porb=227.5687 days)

The prediction for the next nova eruption of TCrB using the method described above has the advantage of providing a precise value, subject only to the uncertainty of the numerical values of the physical quantities involved and the approximations of the underlying theoretical model. Moreover, the result obtained would justify the supposed "delay" observed in the next nova eruption of TCrB.

## 5. *Conclusions*

The symbiotic recurrent nova (SyRNe) TCrb (T-Coronae Borealis) is close to a new nova eruption after the last recorded event in 1946. Careful studies, especially photometric, have led to different predictions for the next nova eruption, taking into account the recurrence times extrapolated from previous eruptions, which an average value about 80 years. Schaefer, in particular, has produced various predictions, among which the one made in 2023 and based on the B&V light curves for the period 1842–2022, stands out, according to which the next nova eruption should occur in 2025.5±1.3 and therefore is currently still valid. Building on Schaefer's recent remarkable work in accurately determining the key physical parameters that drive that drive the dynamics of the TCrB symbiotic system, we have proposed a new semi-empirical method to derive the variations of the nova recurrence time, Trec, and thus obtain a point forecast estimate of the next eruption for the date: 26-Feb-2027 (2027.1545±0.09). This result is currently compatible and consistent with the observed photometric behavior and would also justify the supposed "delay" for the next event of this nova as commented by various authors.

REFERENCES

Livio, Mario, (1988), "Recurrent Novae", J. Mikolajewska et al. (eds.), The Symbiotic Phenomenon, 323-334.

Luna, Gerardo J. M., Sokoloski, J. L., Mukai, Koji, and Kuin N. Paul M. (2020), "Increasing Activity in T CrB Suggests Nova Eruption Is Impending", The Astrophysical Journal Letters, Volume 902, Issue 1, id.L14, 3.

Munari Ulisse, Dallaporta, Sergio, Cherini Giulio, (2016), "The 2015 super-active state of recurrent nova T CrB and the long term evolution after the 1946 outburst", New Astronomy, 47, 7.

Nauenberg, Michael (1972), "Analytic approximations to the mass-radius relation and energy of zer-temperature stars", The Astrophysical Journal, 175:417-430.

Norton, Andrew (2024), "White dwarfs and neutron stars", S384_2, The Open University.

Payne-Gaposchkin, C. (1957), The Galactic Novae (Amsterdam: North Holland).

Schaefer, Bradley E. (2014),"The Recurrent Nova T CrB; Two Discoveries from the 102,000 Magnitude Light Curves from 1855 to 2013 in Johnson B & V", American Astronomical Society, AAS Meeting #223, id.209.01.

Schaefer, Bradley E. (2019), "Predictions for Upcoming Recurrent Novae Eruptions; T CrB in 2023.6±1.0, U Sco in 2020.0±0.7, RS Oph in 2021±6, and more", American Astronomical Society Meeting #234, id. 122.07. Bulletin of the American Astronomical Society, Vol. 51, No. 4.

Schaefer, Bradley E. (2022),"Comprehensive catalogue of the overall best distances and properties of 402 galactic novae", Monthly Notices of the Royal Astronomical Society, Volume 517, Issue 4, pp.6150-6169.

Schaefer, Bradley E. (2023a)," The recurrent nova T CrB had prior eruptions observed near December 1787 and October 1217 AD", Journal for the History of Astronomy, Volume 54, Issue 4, p.436-455.

Schaefer, Bradley E. (2023b),"The B & V light curves for recurrent nova T CrB from 1842-2022, the unique pre- and post-eruption high-states, the complex period changes, and the upcoming eruption in 2025.5 ± 1.3", Monthly Notices of the Royal Astronomical Society, Volume 524, Issue 2, pp.3146-3165.

Schaefer, Bradley E. (2025a), "Orbital Period Changes in Recurrent Nova T Corona Borealis Prove That It Is Not a Type Ia Supernovae Progenitor", The Astrophysical Journal, Volume 991, Issue 1, id.111, 11 pp.

Schaefer, Bradley E. (2025b), "Symbiotic Stars (Including T Corona Borealis) Are Not Immediate Progenitors of Normal Type Ia Supernovae", The Astrophysical Journal, Volume 994, Issue 2, id.187, 10 pp.

Schneider, Jean (2024), "When will the Next T CrB Eruption Occur?", Res. Notes AAS 8 272.

Warner, Brian (1995), Cataclysmic variable stars, Cambridge University Press.

Webbink, Ronald F., Livio, Mario and Truran, James W., (1987) "The nature of the recurrent novae", The Astrophysical Journal, 314:653-672.